\documentclass[usegraphicx,usenatbib]{mn2e}
\usepackage{times}
\usepackage{amsmath}
\usepackage{fixltx2e}

\newcommand{\gmin}{$\gamma_{\rm{min}}$}
\newcommand{\gmax}{$\gamma_{\rm{max}}$}

\begin{document}
\title[ICCMB from Radio Galaxies]{Inverse Compton X-rays from Giant Radio Galaxies at $z\sim$1}
\author[Laskar et al]{T. Laskar$^{1,2}\thanks{Email: tlaskar@cfa.harvard.edu}$,
 A.C. Fabian$^1$, K.M. Blundell$^3$ and M.C. Erlund$^1$\\
1. Institute of Astronomy, Madingley Road, Cambridge CB3~0HA\\
2. Trinity College, University of Cambridge, Cambridge CB2~1TQ\\
3. University of Oxford, Astrophysics, Keble Road, Oxford OX1~3RH}

\maketitle

\begin{abstract}
We report XMM-Newton observations of three FR II radio galaxies at redshifts between 0.85 and 1.34, which show extended diffuse X-ray emission within the radio lobes, likely due to inverse-Compton up-scattering of the cosmic microwave background. Under this assumption, through spectrum-fitting together with archival VLA radio observations, we derive an independent estimate of the magnetic field in the radio lobes of 3C\,469.1 and compare it with the equipartition value. We find concordance between these two estimates as long as the turnover in the energy distribution of the particles occurs at a Lorentz factor in excess of $\sim 250$. We determine the total energy in relativistic particles in the radio emitting lobes of all three sources to range between $3-8\times10^{59}$ erg. The nuclei of these X-ray sources are heavily-absorbed powerful AGN.

\end{abstract}


\section{Introduction}


Emission from inverse-Compton scattered CMB (ICCMB) photons, in the form of diffuse, extended X-rays between the nucleus and radio hot spots, has been detected in a number of radio galaxies at cosmological distances
\citep[e.g.][]{Fabian2003,Blundell2006,Erlund2008,Johnson2007} and at low redshift \citep[e.g.][]{Croston2005}.
ICCMB is a tracer of old, spent synchrotron plasma and its usefulness as a diagnostic of the magnetic field and electron energy distribution in AGN lobes stems from the fact that it is not redshift-dimmed \citep{Schwartz2002}, unlike other sources of continuum X-ray emission.

Together with radio observations, ICCMB X-ray fluxes can be used to constrain the local magnetic field experienced by the
radio-emitting plasma in the lobes of giant FR II \citep{Fanaroff_Riley_1974} radio sources \citep[see e.g.][]{Erlund2006}.
Previous observations of ICCMB X-rays from radio galaxies have found deviations of the observed magnetic field
from predictions based on equipartition arguments \citep[see][and references therein]{Croston2005}.
While this can be resolved if the energy distribution turns over at the low-energy end \citep{Blundell2006} at relatively high Lorentz factors (\gmin$\sim10^3$), estimates of the low-energy turn-over of the electron distribution in all observed radio galaxies and quasars have been largely inconclusive, with no consistent picture amongst the entire population of radio sources having yet emerged.


We present \textit{XMM-Newton} observations of the FR II radio galaxy 3C\,469.1, where we have detected extended X-ray emission,
which is likely ICCMB. We extract and fit the X-ray spectrum for this source and show the photon spectral index, $\Gamma$
to lie between $0.9$ and $2.1$, consistent with the radio synchrotron spectral index of $1.97$ determined from NED\footnote{http://nedwww.ipac.caltech.edu} archival radio photometry.
We use the calculated unabsorbed X-ray luminosity along with a new look at archival 1.5 GHz VLA data for this source to constrain the magnetic field, \textit{without} assumptions of equipartition. 
Finally, we present X-ray observations of two more giant radio galaxies, MRC\,2216-206 and MRC\,0947-249,
where ICCMB emission seems to be also apparent.

3C\,469.1, MRC\,2216-206 and MRC\,0947-249 are located at redshifts of 1.336, 1.148 and 0.854 respectively, corresponding to linear scales of 8.6 kpc/\arcsec, 8.4 kpc/\arcsec and 7.8 kpc/\arcsec\ in our adopted cosmology of $H_0=70\ \rm{km}\ \rm{s}^{-1}\rm{Mpc}^{-1}$ and $\rm{\Omega_\Lambda}=0.73$ and assuming a flat universe.

\section{Observations and Data Analysis}
\subsection{X-ray and radio imaging}
Our \textit{XMM-Newton} observations of 3C\,469.1 were taken on 2008 September 29 and consist of 17.4 ks of EPIC-PN and 21.3 ks of EPIC-MOS data. Although the MOS data give roughly consistent results, they are much noisier and are not presented here. The EPIC-PN data were reduced using the standard XMM-SAS pipeline tasks\ EPCHAIN to give $443\pm21$ counts (of which $110\pm4$ are background) in the $0.3$-$10$ keV band.


\begin{figure*}
\begin{tabular}{cccc}
\centering
 \includegraphics[width=0.35\columnwidth]{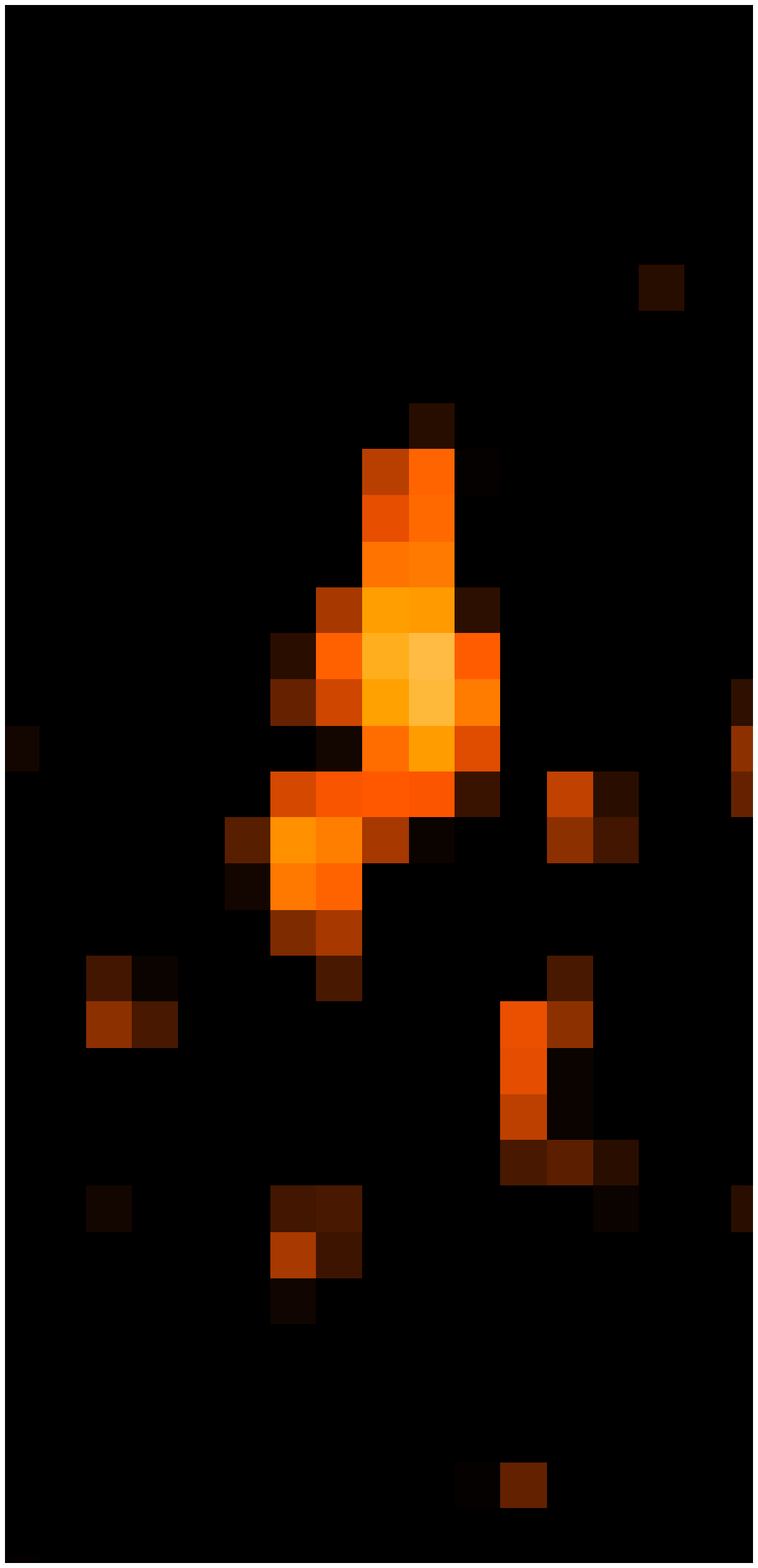} &
 \includegraphics[width=0.35\columnwidth]{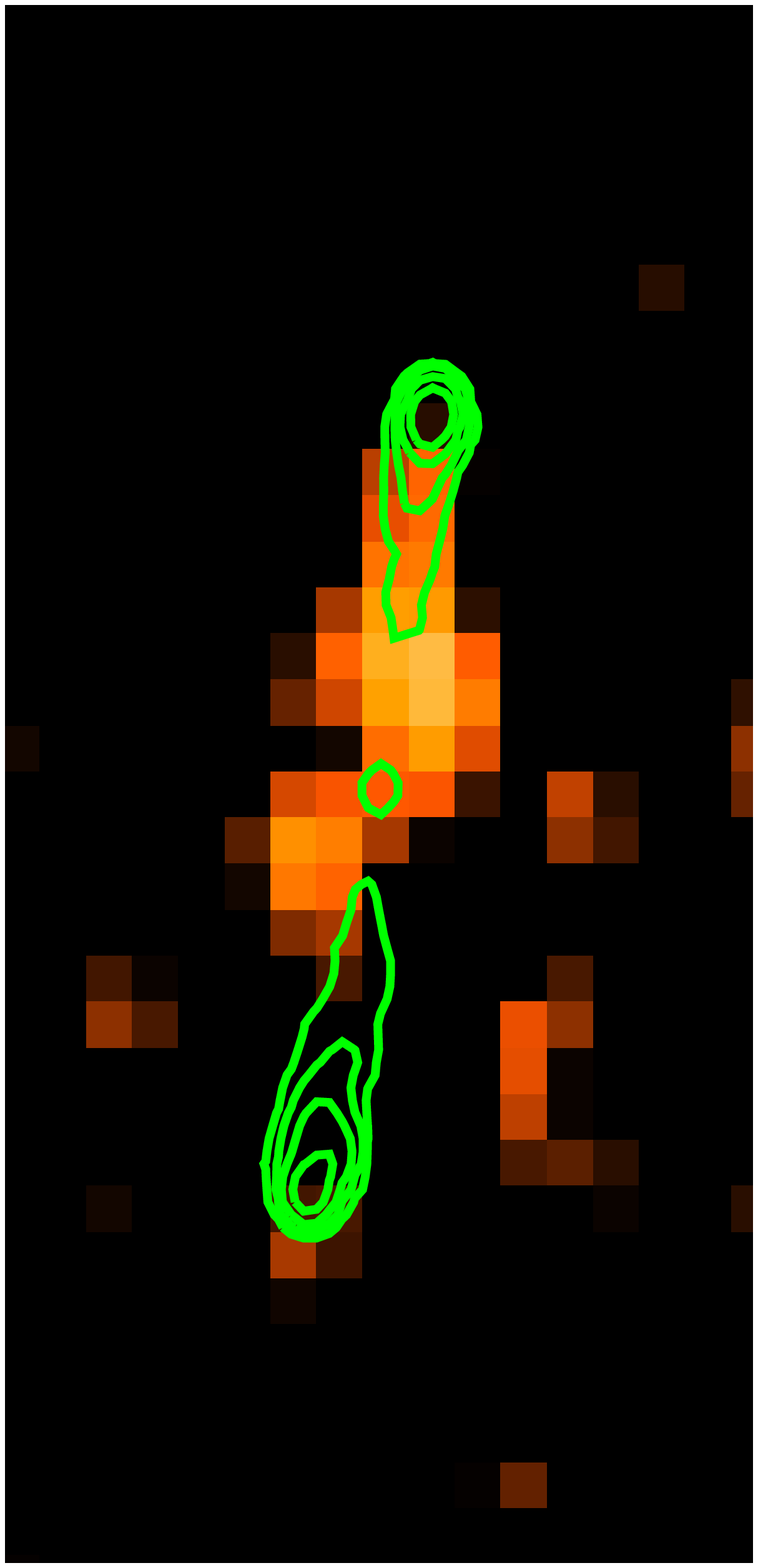} &
 \includegraphics[width=0.35\columnwidth]{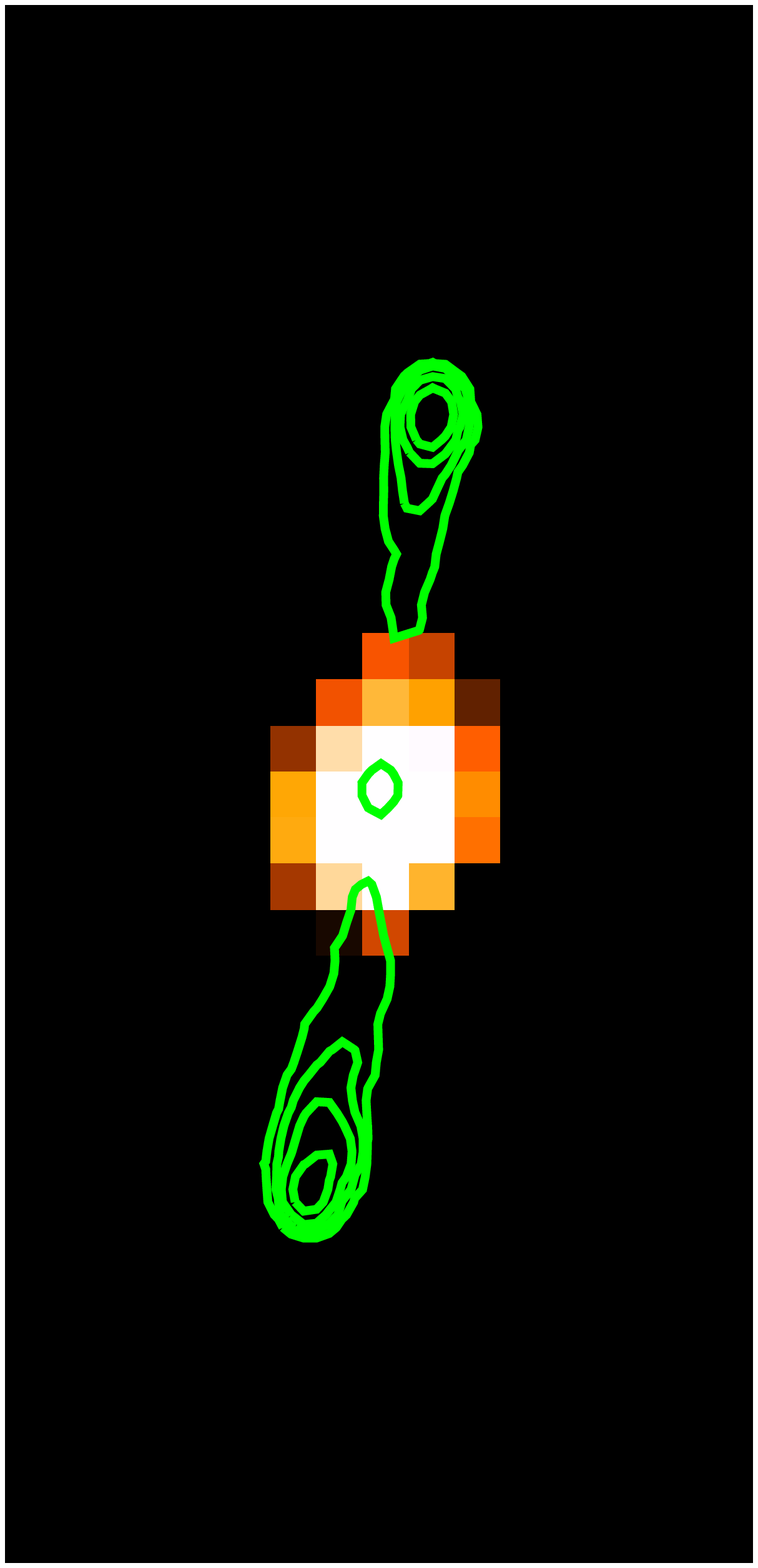} &
 \includegraphics[width=0.35\columnwidth]{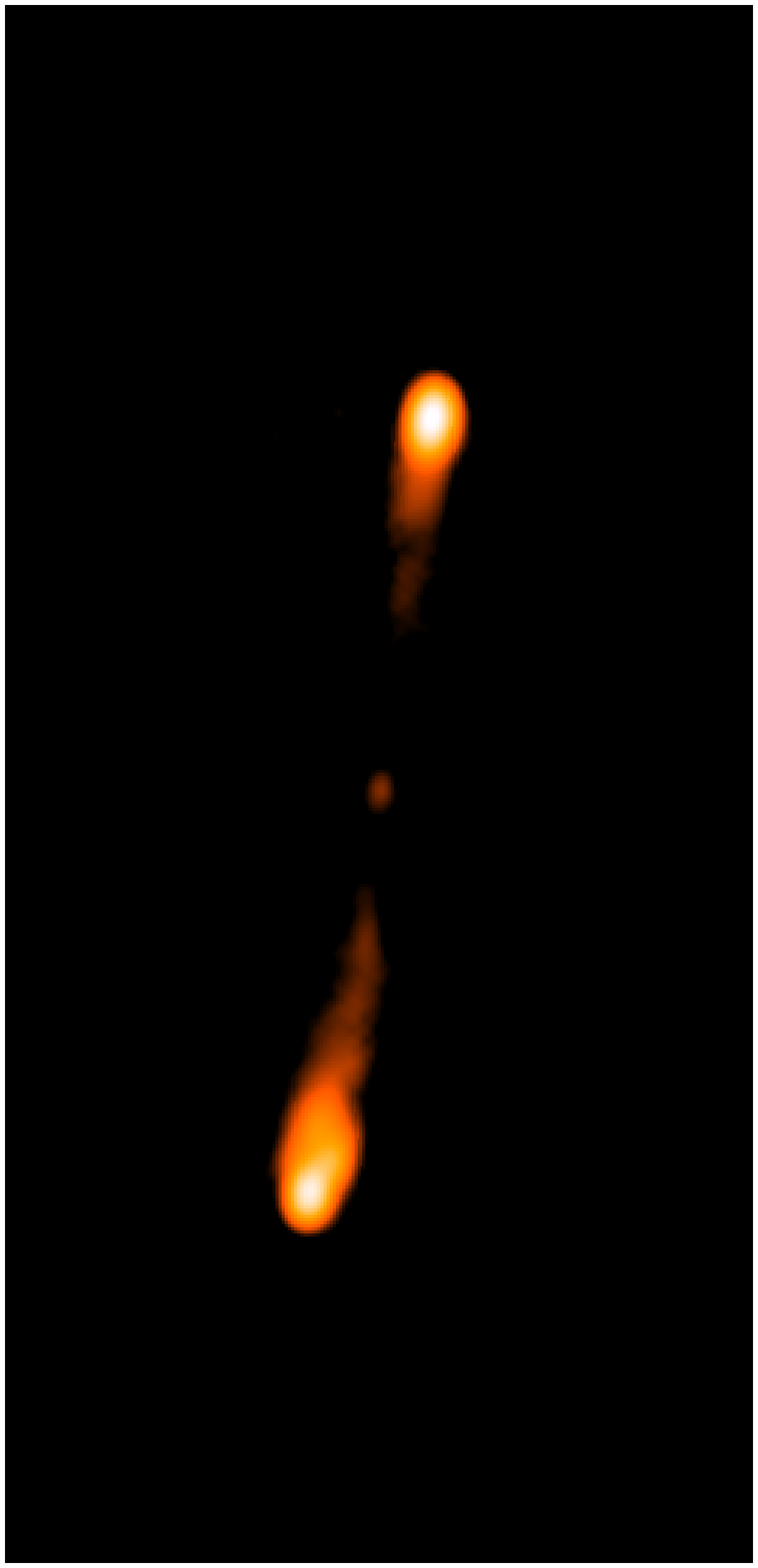}\\
\end{tabular}
\caption{The panels show (from left to right) the XMM pn image of 3C\,469.1 below 1.2 keV, the same with 1.4 GHz radio contours overlaid, the pn image above 2 keV with the same radio contours, and the 1.4 GHz radio image. The pixels in the X-ray images are 4.35 arcsec across. Each image is 150 arcsec (1.29 Mpc) high. All X-ray images (except Fig. \ref{mrc2216_coresplit}) 
have been convolved with a Gaussian with FWHM of 2 pixels, which is roughly the FWHM of the PSF.
}  \label{coresplit}
\end{figure*}

The source clearly shows extended emission at energies below $1.2$ keV (Fig. \ref{coresplit}). We estimate the detection significance of the entire extended emission to be $10$-$\sigma$ and of the northernmost region of the quasi-linear structure to be $5$-$\sigma$. Above 2 keV, a central core remains, which is consistent with being a point source. 
\textbf{The extended linear morphology of the soft X-ray emission precludes it being due to hot cluster gas.} 


Radio data from all three targets came from the VLA archive. In the case of 3C\,469.1, we used data from the A, B and C configurations (originally published by \citealt{Kharb2008}) to sample all the structure; for our other two targets only B-configuration data were available. Interference was excised from the data and 3C48 was used to set the absolute flux calibration; a number of rounds of phase self-calibration were performed, before a final iteration of phase and amplitude self-calibration. The final self-calibrated 1.5 GHz radio image for 3C\,469.1 is shown in Figure \ref{coresplit}. \textbf{The cospatial nature of the radio synchrotron emission compared with the soft X-rays from this source suggests a fundamental connection between the emitting particle populations within the lobes.}

The flux density of the extended radio emission, after subtracting the contribution from the hotspots, is about 300 mJy, with the North and South radio lobes contributing $0.256$ Jy and $0.046$ Jy, respectively. We note that the southern lobe is not detected in the soft X-ray image. A possible explanation for the extended emission is Inverse-Compton scattering of infrared photons from the host galaxy \citep{Brunetti1997} into the line of sight by plasma within the lobes, which would produce more intense X-rays in the lobe directed away from the observer. However, in order for the energy density of the nuclear photons to be comparable to that of the microwave background 200 kpc into the lobes, the nucleus would be required to have an infrared luminosity of $10^{48}$ erg s$^{-1}$, which is implausible. Therefore, it is more likely that ICCMB is responsible for the observed X-ray flux. The discrepancy in the flux from the two lobes remains unresolved.

\subsection{X-ray spectral analysis of 3C\,469.1}
\begin{figure}
\centering
 \includegraphics[width=0.69\columnwidth,angle=-90]{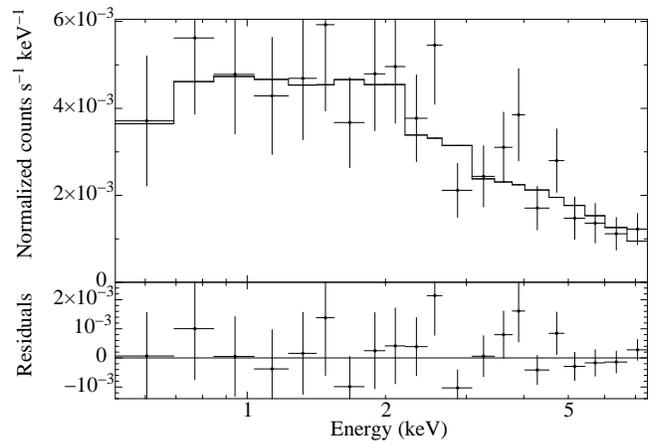}
\caption{XMM pn spectrum of 3C\,469.1. The model fitted is shown in Figure \ref{model} and is
described by equation \ref{modelequation} with the best fit parameters as shown in Table \ref{bestfit}.}
 \label{spectrum}
\end{figure}

The number of counts was insufficient to perform spectral fitting of the lobe and core regions separately (an attempt at fitting only the extended X-ray emission towards the North with a power law yielded a poorly-constrained photon index of $\Gamma=3.2^{+1.2}_{-1.6}$).
Therefore spectra of the entire source were extracted, with regions on the same chip selected as background. The spectrum was binned to 20 counts bin$^{-1}$ using GRPPHA and fitted in XSPEC v12.5.0 (Fig. \ref{spectrum}).

The model used was a set of power laws in photon energy for both the extended and core emission, together absorbed photoelectrically by neutral hydrogen within our Galaxy,
with allowance for additional soft-photon absorption within the host galaxy itself.
Due to the poor constraints from the data, we assumed the two power law indices to be the same.
Mathematically, the model of the X-ray flux density as a function of the photon energy, $E$:
\begin{equation}
 M(E) = e^{-N^1_{\rm{H}} \sigma(E)}
(K_1 E^{-\Gamma_1} + e^{-N^2_{\rm{H}} \sigma(E\cdot(1+z))} K_2 E^{-\Gamma_2}).
\label{modelequation}
\end{equation}

\begin{figure}
 \includegraphics[width=0.65\columnwidth,angle=-90]{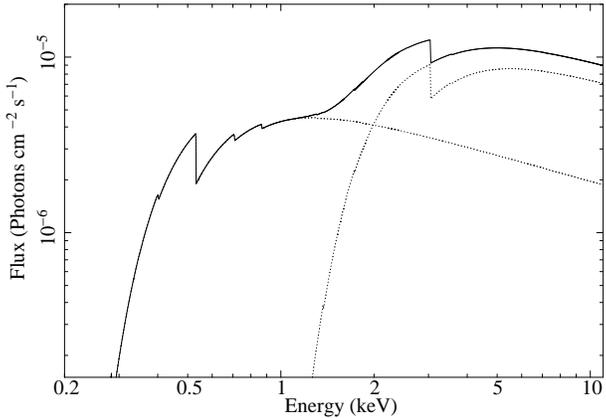}
\caption{Best-fit model for XMM pn data for 3C\,469.1. The model used (equation \ref{modelequation})
is two absorbed power laws with the same photon index, with one power-law component further attenuated by
an absorbing column (estimated at 
$29.4_{-17.6}^{+18.3}\times 10^{-22}$ cm$^{-2}$ at 90 per cent confidence) within the host galaxy.
}
 \label{model}
\end{figure}

\noindent Here $K_1$ and $K_2$ are the normalisations of the emission 
at 1 keV from the lobes  and the core, respectively, $\Gamma_1$ and $\Gamma_2$ are the corresponding photon indices (set equal),  $N^1_{\rm{H}}$ is the absorbing column  within our Galaxy along the line of sight to the source ($1.38\times10^{21}$ cm$^{-2}$; \citealt{Kalberla2005}),  $N^2_{\rm{H}}$ is the extra absorbing column within the host galaxy, $\sigma(E)$ is the photo-electric cross-section  (ignoring Thomson scattering) and $z = 1.336$ is the redshift. The best fit model ($\chi^2 = 13.9$ for  17 d.o.f.) is shown in  Fig. \ref{model} and the best fit values for this model are given in Table \ref{bestfit}.

\begin{table}
\begin{tabular}{l l l}
\hline
Parameter & Value & Nominal error\\ & & (90 per cent confidence)\\
\hline

$\Gamma_1 \equiv \Gamma_2$  & 1.5 & 0.6\\
$N^2_{\rm{H}}$           & $29$ & $\pm18$\\
$K_1$   & $6.3$  & ${}^{+1.5}_{-1.7}$\\ 
$K_2$   & $24$  & ${}^{+48}_{-20}$\\

\hline
\end{tabular}
\caption{The constraints on the parameters for the best-fitting model (equation \ref{modelequation}).
$N^2_{\rm{H}}$ is the extra absorbing column within the host in $10^{22}$ cm$^{-2}$.
$K_1$ and $K_2$ are the normalisations of the emission (in $10^{-6}$ keV$^{-1}$ cm$^{-2}$ s$^{-1}$) at 1 keV
from the lobes and the core, respectively.\label{bestfit}}
\end{table}

\begin{figure}
\includegraphics[width=0.69\columnwidth,angle=-90]{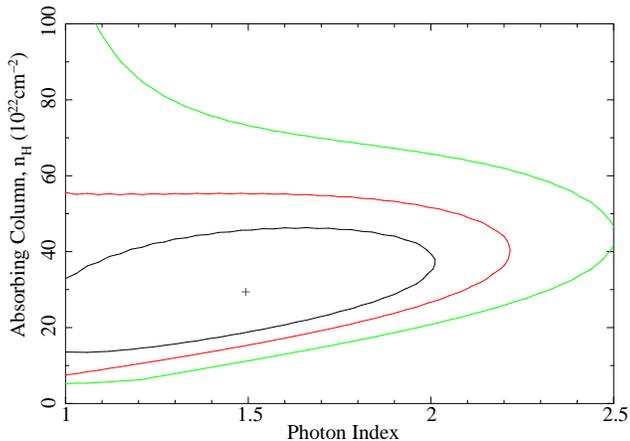}
 \caption{$\chi^2$ contours at 68 per cent (innermost, black), 90 per cent (middle, red) and 99 per cent (outermost, green) confidence between the absorbing column in the host galaxy, $N^2_{\rm{H}}$ and the power-law photon index, $\Gamma$ for the best
fit model for the 3C\,469.1 spectrum.}
 \label{contourplot_2_4}
\end{figure}

The best fit model calls for a high absorption column in the nuclear component. A plot of the $\chi^2$ contours of $N^2_{\rm{H}}$ versus $\Gamma$ (Fig. \ref{contourplot_2_4}) shows the extra absorption to be statistically significant.
After correcting for Galactic absorption, the total luminosity in the rest frame of the source in the $2$-$10$ keV band is
 $5.75\times10^{44}$ erg s$^{-1}$, of which $3.25\times10^{44}$ erg s$^{-1}$ comes from the (absorbed) core component. The remaining luminosity of $2.5\times10^{44}$ erg s$^{-1}$, corresponding to a flux density of
$2.8\times10^{-14}$ erg s$^{-1}$ cm$^{-2}$ ($1.15$ nJy for an X-ray bandwidth of $\sim$ 8 keV)
 at $z=1.336$ from the unabsorbed (lobe) component, is comparable to the power in the X-ray emitting lobes of the giant radio galaxy, 6C0905+39  ($1.5\times10^{44}$ erg s$^{-1}$; \citealt{Erlund2008}).
After correcting for both Galactic and host absorption, the $2-10$ keV rest-frame luminosity of the nucleus is $9.8\times10^{44}$ erg s$^{-1}$.

There are 22 photometric data points from 37 MHz to 14.9 GHz for 3C\,469.1 in the NASA Extragalactic Database. A log-log plot of flux density versus frequency yields a straight line with a slope of $-0.97$ and a correlation coefficient of $r^2 = 0.98$. This implies an electron energy distribution proportional to $E^{-2.94}$, which is considerably steeper than the distribution ($E^{-(2\Gamma-1)}\propto E^{-2.0}$) implied by the X-ray spectra. However, our X-ray data only poorly constrain the photon spectral index and a value of $\Gamma=1.97$, as expected from the radio photometry, is entirely within the ($2\sigma$) uncertainty (Fig. \ref{contourplot_2_4}). Also, this fit does not take into account the core photon index separately, which may have a flatter spectrum than the lobes, forcing the overall best fit to lower values of $\Gamma$. Further, taking into account the noisy MOS data pushes up the value of $\Gamma$ from $1.5$ to $1.7$. In what follows, we shall assume an electron energy distribution of $N_{\rm{e}}(E)\propto E^{-2.94}$ ($\Gamma=1.97$) for this object.

\section{Determination of Magnetic Field}
\label{sec:determinationofB}
For an electron population with a power-law distribution of energies, $N_{\rm{e}}(E)\propto E^{-n}$
the synchrotron photons also follow a power-law spectrum with a spectral index, $\alpha=(n-1)/2$.
The ratio of the flux density from inverse-Compton emission at $\nu_{\rm{c}}$ to the synchrotron flux density at $\nu_{\rm{s}}$
is given by \citep[see, e.g.][]{Tucker1977}
\begin{equation}
\begin{split}
\frac{F_{\rm{c}}}{F_{\rm{s}}} &= 2.47\times10^{-19}(5.23\times10^3)^\alpha\left(\frac{T}{\mbox{1 K}}\right)^{3+\alpha}\times\\
&\qquad\frac{b(n)}{a(n)}\left(\frac{B}{\mbox{1 Gauss}}\right)^{-(\alpha+1)}
\left(\frac{\nu_{\rm{c}}}{\nu_{\rm{s}}}\right)^{-\alpha},
\end{split}
\label{getB}
\end{equation}
where $B$ is the magnetic field strength in the radio-emitting lobes and $T$ is the temperature of the CMB at the redshift of the source, and the constants $a(n)$ and $b(n)$ have been tabulated by \citet{Ginzburg1965} and \citet{Tucker1977}, respectively.

Taking the flux density of the inverse-Compton emission to be the X-ray flux density from $2-10$ keV (rest frame), 
the flux density of the synchrotron emission to be 300 mJy at
1505 MHz, $\alpha = 0.97$, and $h\nu_{\rm{c}}\approx3$ keV yields a magnetic field of 12.5\,$\mu$G. For comparison, taking $\alpha=0.7$ in the 
above formula yields $B=6.8$\,$\mu$G, keeping all other values the same. It should be noted that this analysis requires the observed X-ray and radio fluxes to be co-spatial. If the X-ray emission is more extended than the radio, then one must use in the above formulae only that part of the X-ray flux which overlays the radio and vice-versa.


One can also determine the magnetic field by assuming the energy density in it to be the same as that of the electron population
\citep[see][eqn. 18.73]{LongairHEA2}. Taking the radio emitting lobes of 3C\,469.1 to be a cylindrical region 640 kpc long and 140 kpc in diameter and assuming a minimum Lorentz factor of the electron population, $\gamma_{\rm{min}} = 10^3$ gives an equipartition magnetic field of $9.0$\,$\mu$G. While this value is consistent with that determined from the X-ray and radio flux densities, the equipartition field depends on the volume of the region ($B\propto V^{-3.97}$ for $\alpha=0.97$), which is difficult to determine with the low spatial resolution of XMM and given the low number of counts. With the assumed volume stated above, we must have $\gamma_{\rm{min}} > 250$ for $\alpha\leq0.97$ in order for the source to be in equipartition and to simultaneously explain the ICCMB X-ray flux.
\textbf{The volume emissivity in the ICCMB X-rays,
\begin{equation}
 j(\nu)= 4.19\times10^{-40} T^3(z) b(n) N\left(\frac{2.1\times10^{10} }{\nu} T(z)\right)^{(n-1)/2}
\label{KE1}
\end{equation}
integrated over rest frame energies from 2-10 keV and equated to the X-ray luminosity of the lobes, yields the number density of relativistic electrons to be $N/V=2.2\times10^{-3}\rm{m}^{-3}$.} Integrating the electron energies over a power-law distribution with index, $n$ between Lorentz factors \gmin\ and \gmax\ now allows for the determination of the total energy of the electrons responsible for the ICCMB.
From the relation
\begin{equation}
 \gamma \sim 1.89{\sqrt{1+z}}\left(\frac{\nu_{\rm{R}}}{1\,\rm{GHz}}\right)^{1/2}\left(\frac{B}{1\,\mu\rm{G}}\right)^{-1/2}10^4,
\label{gammacalcr}
\end{equation}
the synchrotron emission at 14.9 GHz \citep{Laing1980}
corresponds to Lorentz factors of $3.2\times10^4$. Taking \gmax\ to be this value and assuming the electron energy spectrum to extend down to \gmin\ $= 10^3$, the total energy in the plasma between \gmin\ and \gmax\ is
$8.1\times10^{59}$ erg
, which is comparable to that found by \citet{Erlund2008} in
the giant radio galaxy 6C0905+39 and by \citet{Fabian2009} for HDF130 in the Chandra Deep Field North. \textbf{The combined pressure produced by relativistic electrons and the magnetic field is $P/\rm{k_B}=5.2\times10^{4}$ Kcm$^{-3}$, which is comparable to that near the centre of a group of galaxies, but much larger than that of the IGM ($\sim$ 0.3 K/cm$^{-3}$). Therefore whether the lobes continue to expand depends on the environment.} 

\section{MRC Sources}
\begin{table*}
\begin{minipage}{155mm}
\begin{tabular}{c c c c c c c c c c}
\hline
Source & RA & Dec & $z$ & countrate & $N_{\rm{H}}$ & Unabsorbed & L$_{44}$ & $\mathcal{E}_{\rm{e}}$\\
       & (J2000) & (J2000) &  & [cts/s]   & [$10^{21}$ cm$^{-2}$] & Flux & \\
\hline
3C\, 469.1    & 23h55m23.3s & +79d55m20s &1.336& $1.93\times10^{-2}$ & $1.38$ & $3.58$ & 2.56 & 8.1\\
MRC\,2216-206 & 22h19m44.2s & -20d21m31s &1.148& $2.09\times10^{-2}$ & $0.25$ & $2.93$ & 1.63 & 5.7\\
MRC\,0947-249 & 09h49m52.9s & -25d11m40s &0.854& $4.50\times10^{-3}$ & $0.46$ & $1.22$ & $0.45$ & 2.9\\
\hline
\end{tabular}
\caption{Observed X-ray luminosities for the sources in this paper.
$N_{\rm{H}}$ is the Galactic neutral hydrogen column density along the line of sight, calculated using PIMMS.
Columns 7 and 8 show the observed X-ray flux in $10^{-14}$ erg cm$^{-2}$ s$^{-1}$ and the 
X-ray luminosity in $10^{44}$ erg s$^{-1}$ ($2-10$ keV, rest frame) respectively, both corrected for Galactic absorption. Column 9 shows the total energy in relativistic electrons in $10^{59}$ erg, estimated using equation (\ref{KE1}) for 3C\,469.1 and equation (\ref{KE2}) for the MRC sources. For 3C\,469.1 and MRC\,2216 the quoted values of flux, luminosity and energy are for the lobe component as determined from spectrum-fitting, while for MRC\,0947 they have been calculated from the observed soft ($0.3-1.5$ keV) X-ray count rate of $4.6\times10^{-3}$ s$^{-1}$ using PIMMS. 
\label{mrcfluxes}
}
\end{minipage}
\end{table*}



\begin{figure}
\centering
\begin{tabular}{c c}
 \includegraphics[width=0.45\columnwidth]{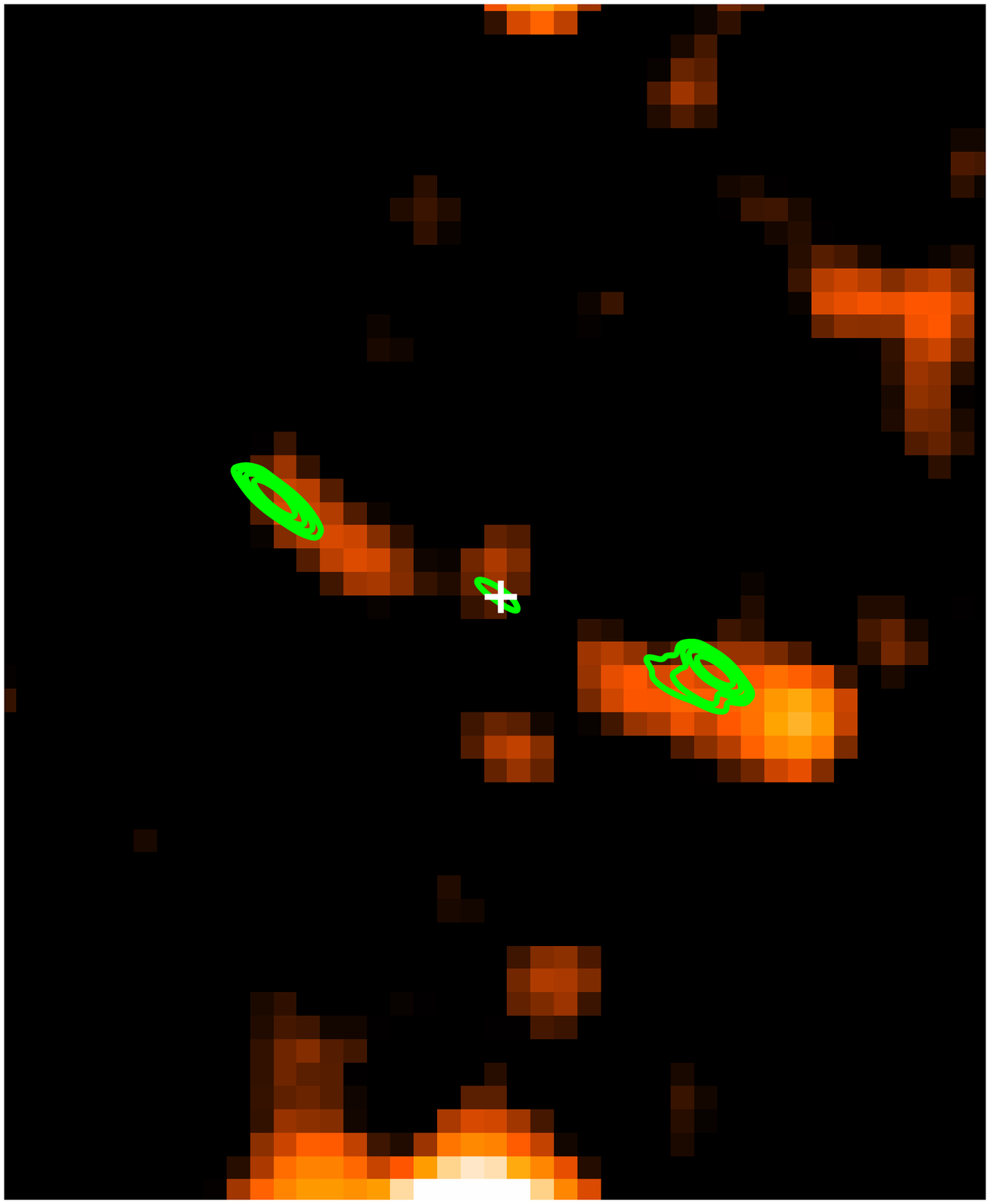} &
 \includegraphics[width=0.45\columnwidth]{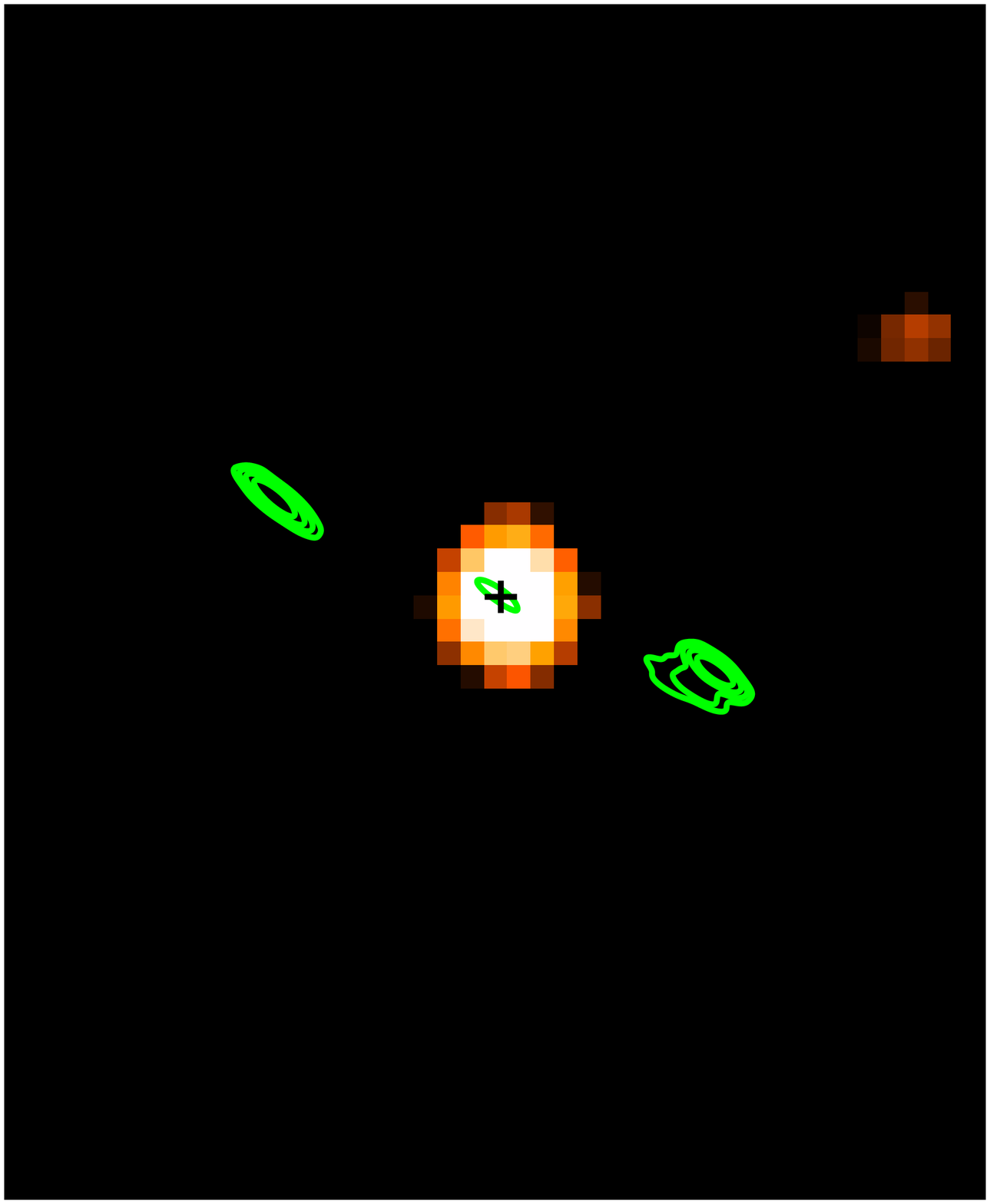} \\
\end{tabular}
\caption{XMM-pn image MRC\,2216-206 from $0.3-1.5$ keV (left panel) and $1.5-8.0$ keV (from right panel), 
with 4.86 GHz radio contours overlaid. Each panel is 222 arcsec high and 183 arcsec wide. The X-ray images have been convolved with a Gaussian with FWHM of 3 pixels.}
\label{mrc2216_coresplit}
\end{figure}

\begin{figure}
\centering
 \begin{tabular}{c c}
 \includegraphics[width=0.45\columnwidth]{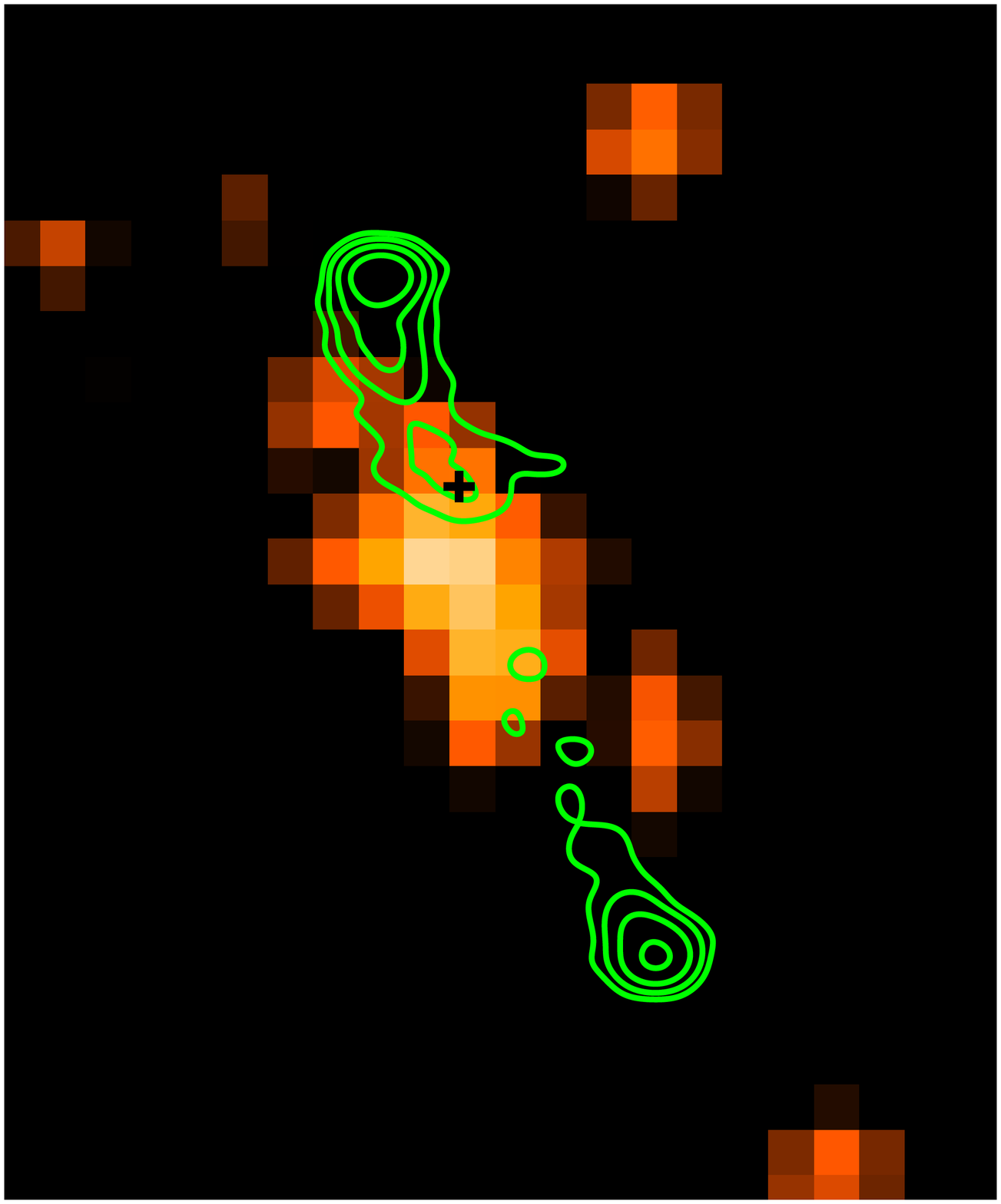} &
 \includegraphics[width=0.45\columnwidth]{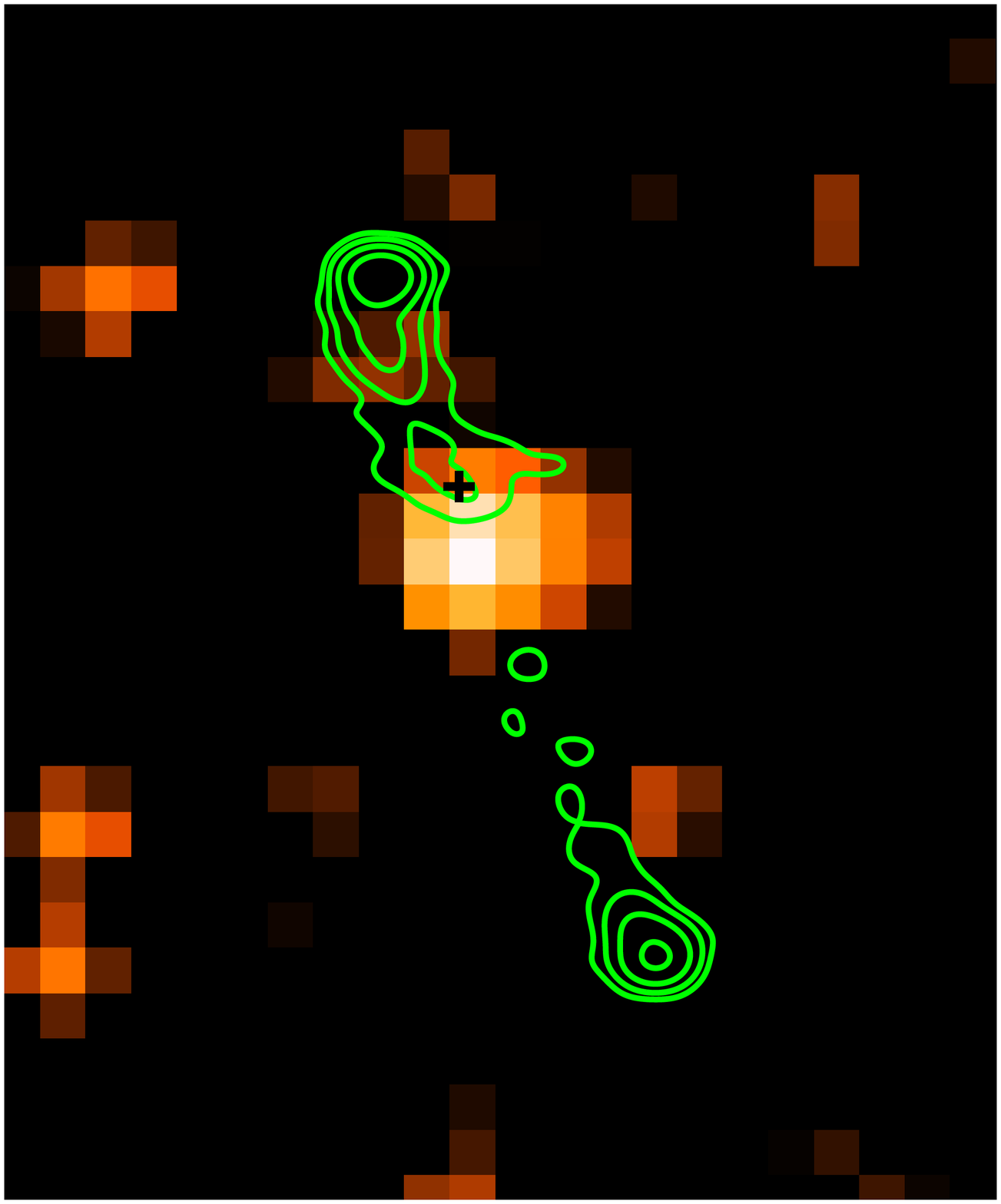} \\
\end{tabular}
\caption{XMM pn image of MRC\,0947-249 from $0.3-1.5$ keV (left panel) and $1.5-8.0$ keV (right panel), 
with 4.86 GHz radio contours overlaid. Each panel is 114 arcsec high and 95 arcsec wide.}
\label{mrc0947_coresplit}
\end{figure}

We also obtained observations of giant radio galaxies MRC\,2216-206 ($z=1.148$, linear size 730 kpc \citep{Kapahi1998}) and MRC\,0947-249 ($z=0.854$, linear size 530 kpc \citep{Kapahi1998}) with XMM-Newton on 2008 May 03 and May 10, respectively, with corresponding exposure times of
22.6 ks (8.8 ks)\footnote{Good time remaining after filtering to remove flares and taking dead-time intervals into account}
and 17.0 ks (10 ks). In the $0.3-10.0$ keV band, there are $361\pm19$ counts (of which $181\pm6$ are background, with $47\pm19$ soft\footnote{$0.3-1.5$ keV \label{defsoft}} photons) for MRC\,2216-206 and $195\pm14$ counts (of which $126\pm6$ are background, with $46\pm11$ soft photons) for MRC\,0947-249. Whereas at higher energies the X-ray images are more consistent with being point-like sources near the location of the corresponding optical host galaxy, indicated by crosses (Fig. \ref{mrc2216_coresplit} and \ref{mrc0947_coresplit}), both sources show extended soft X-ray emission below 1.5 keV, which is plausibly due to ICCMB.

We note that the soft emission in MRC\,2216-206 appears to extend beyond the location of the South-East radio hotspot. This could indicate an earlier outburst resembling the double-double radio galaxy phenomenon \citep{Schoenmakers2000} but with the outer/older lobes only being detectable via ICCMB. However this extra emission, which consists of 30 counts in the $0.3-1.5$ keV band, may also be due to a background AGN (at the corresponding flux level of $6\times10^{-15}$erg s$^{-1}$ cm$^{-2}$ in the $0.5-2.0$ keV band for a photon index of $\Gamma=2.0$ for these 30 counts, we estimate a 30 per cent probability of a coincident background source in the regions of interest around our three sources using the source counts of \citealt{Hasinger2001}).
We did not include the region beyond the radio hotspots in our analysis.

\begin{figure}
\centering
 \includegraphics[width=0.74\columnwidth,angle=-90]{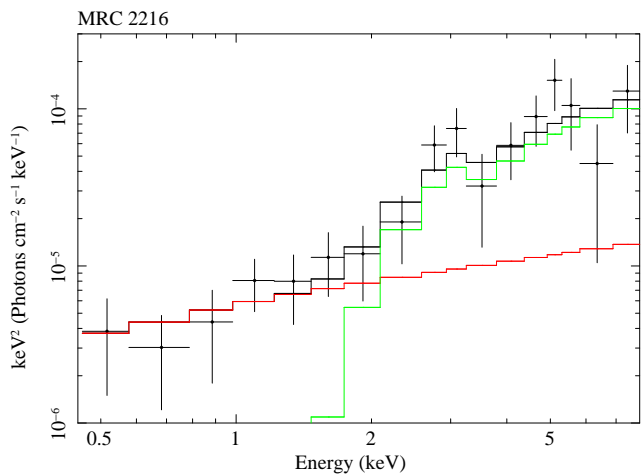}
\caption{Unfolded XMM pn spectrum of MRC\,2216-206. The model fitted is described by equation \ref{modelequation}. Note that the y-axis is equivalent to $\nu F_{\nu}$. The solid red line depicts the
extended ICCMB component and the green solid line the absorbed nuclear component.} 
 \label{MRCspectrum}
\end{figure}

Spectral analysis of MRC\,2216-206 was carried out exactly as for 3C\,469.1, since a single-component power
law fit gave an unphysically hard photon index of $\Gamma=0.5\pm0.35$. The two-component model yielded an extra host absorption of $3.8\pm1.8\times 10^{23}$ cm$^{-2}$ and a best fit energy index, $\Gamma=1.6^{+0.5}_{-0.4}$, consistent with a radio synchrotron index of $\alpha=1$ derived from NED archival radio photometry.
The unabsorbed $2-10$ keV luminosity of the nucleus is $1.3\times10^{45}$ erg s$^{-1}$.
The unfolded spectrum, along with the best fit model and residuals is shown in Fig. \ref{MRCspectrum}.

If we assume that the observed X-rays are inverse-Compton CMB photons, we can follow equation 4 of \citet{Erlund2006} to estimate the total energy in relativistic particles from the X-ray flux alone:
\begin{equation}
 \mathcal{E}_{\rm{e}} = \frac{3}{4}\frac{L_{44}}{\gamma_{\rm{e}}(1+z)^4}10^{64} \rm{~erg},
\label{KE2}
\end{equation}
where $L_{44}$ is the X-ray luminosity in $10^{44}$ erg s$^{-1}$ and $\gamma_{\rm{e}}$ is the typical Lorentz factor of the electrons up-scattering the CMB to X-rays. Setting  $\gamma_{\rm{e}}=10^3$ gives $\mathcal{E}_{\rm{e}} = 5.7\times 10^{59}$ erg for MRC\,2216-206. This is a lower limit to the total particle energy in the lobes of this object.

There were insufficient counts to fit the spectrum of MRC\,0947-249. Instead, the soft photon count rate ($4.6\times10^{-3}$ s$^{-1}$) was used along with a power-law model and an assumed photon energy index of $\Gamma=2.0$ to estimate the flux of the extended emission with PIMMS. The luminosity and total energy in relativistic particles thus determined are given in Table \ref{mrcfluxes}, together with the corresponding values for 3C\,469.1 and MRC\,2216-206 determined from spectrum-fitting.
\section{Conclusions}

We have detected extended X-ray emission in the giant FRII radio galaxy, 3C\,469.1, which extends up to the northern radio hotspot. 
The spectrum is well fit ($\chi^2 = 13.9$ for 17 d.o.f.) by a power law with a photon index consistent with the radio spectral index. Energy equipartition arguments agree with the magnetic field estimates made for 3C\,469.1 based on the X-ray flux inferred to be ICCMB, provided \gmin\ is in excess of 250.

Further, we have presented XMM images of two more radio galaxies at cosmological distances, which show extended X-ray emission attributable to inverse-Compton CMB.
The nuclei of all these sources have quasar-like luminosities and appear to be heavily absorbed powerful AGN, while the radio-emitting lobes have relativistic particles with similar total energies, ranging between 3 and $6\times10^{59}$ erg. These values are enhanced if there is significant energy in protons. 

It appears that ICCMB X-ray emission is indeed observable from giant radio galaxies at redshift $z\sim1$.
However, the cospatial nature of the X-rays with radio observations remains an important caveat in the determination of the magnetic field. Further observations of these sources with the finer PSF of \textit{Chandra} would help resolve this issue and observations with greater signal-to-noise would provide better constraints on the X-ray spectral index, which is essential for accurate determinations of \gmin\ and the establishment of the nature of the low-energy turn-over in aged AGN lobes.


\section*{Acknowledgements}
ACF and KMB thank The Royal Society for support. This work was supported by the Cavendish Laboratory, Cambridge and a scholarship from the Jawaharlal Nehru Memorial Trust and the Cambridge Commonwealth Trust.

\bibliography{iccmb2009}
\bibliographystyle{mn2e}

\end{document}